\documentclass[twocolumn,english,aps, pre]{revtex4}
\usepackage[T1]{fontenc}
\usepackage[latin9]{inputenc}
\usepackage{textcomp}
\usepackage{amstext}
\usepackage{graphicx}



\usepackage{babel}
\begin{document}

\title{Electric properties of supercooled water contained in cylindrical nanopores}

\author{G.S. Bordonskiy, A.O. Orlov}

\affiliation{Institute of Natural Resources, Ecology and Cryology SB RAS, 16 Nedorezova,
p/b 521, 672014 Chita, Russia}

\email{lgc255@mail.ru}

\begin{abstract}
The paper provides data on measuring electrical properties of supercooled
water in nanoporous silica MCM-41 with $3.5\,nm$ diameter cylindrical pores,
using the methods of dielectric spectroscopy and measuring proper electrical
fluctuations at low frequencies. Occurrence of non-linear media properties
at the temperatures below $-35^{\circ}\,C$ was determined, which was revealed
in the form of registered cell capacity dependence on voltage amplitude
in it, as well as noise increase close to $-40^{\circ}\,C$. The effects
observed are supposed to be related to the earlier predicted ferroelectric
phase transition. 
\end{abstract}

\maketitle

\section*{Introduction}

One of the methods of studying the properties of supercooled water is the
method of dielectric spectroscopy of pore materials containing entrapped
water. It is widely used in the frequency range $0.01\,Hz$ to $10\,MHz$ \cite{Frunza2004studies,Sliwinska2008melting}.
In some cases the measurements are taken over the frequency of $1\,GHz$ \cite{Frunza2004studies},
and measurements of some parameters - over the frequency of $22\,GHz$ \cite{Bordonskiy2012structural}.

When combining the dielectric spectroscopy method with the methods of
neutron and X-ray diffraction and scanning differential spectroscopy,
some unique data on structural water transformations in nanoporous
media were obtained. For instance, occurrence of cubic ice Ic at the
temperatures below $-33^{\circ}\,C$ was experimentally determined \cite{Sliwinska2008melting,Johari2009}. It was
reported about detection of ferroelectric phase transition close to
$-40^{\circ}\,C$ in supercooled water in MCM-41 silica pores which look like
$3.5\,nm$ diameter ordered cylinders \cite{Bordonskiy2011study,Bordonskiy2012study,FMBO2011}. The phase transition in
supercooled water at the temperatures $-45\,\div\,-37^{\circ}\,C$ was theoretically
discussed in \cite{A1983,MF2011}.

Detection of nonlinear electrical properties in porous moistened media
requires the dielectric spectroscopy method correction. When studying
ferroelectric phase or phase transition areas, medium response to
external electric field can depend on amplitude, frequency and some
other factors in a complicated way. 

The purpose of the present study was to search for signs of supercooled
water phase transition at the temperature close to $-40^{\circ}\,C$, from
paraelectric to ferroelectric phase, for confirmation of the research
results \cite{Bordonskiy2011study,Bordonskiy2012study,FMBO2011,A1983,MF2011}. This transition must be accompanied by the change in
medium electric polarization, so methods of electric measurements should
be used for its identification. Moreover, we had the task of dielectric
spectroscopy method modification to study nonlinear ferroelectric media.

\section*{Methods of research}

Improving the dielectric spectroscopy technique, we used the idea of a
nonlinear electric circuit and its reaction to an external signal which
depends on amplitude and frequency of a test signal. However, some
complex parametric phenomena occur, and negative electrical resistance
may appear \cite{Kh2002}. Electromagnetic oscillation can appear in this system
causing total measurement results distortion in the standard dielectric
spectroscopy technique \cite{BM2005}. Therefore, the research task was to detect
the negative resistance and conditions of its occurrence in moist
nanoporous silicates. Another aspect of measurement modification
consisted in measuring capacity of a cell containing the medium studied
with different voltage of its electric field. Occurrence of powerful
electric polarization and increase of dielectric constant must have
caused spasmodic rise in cell capacity, when cooling below $-40^{\circ}\,C$ \cite{Bordonskiy2011study,Bordonskiy2012study,FMBO2011}.
However, it wasn't reported about any findings on this effect in the
studies that had been done by other researchers previously. We were
to find out the causes for this disagreement.

A method for measuring electric noise EMF used for studying properties
of ferroelectrics is also applied during the research \cite{Bed2011}.
In particular, this method uses intrinsic thermal noises for
registration of changes in absolute sample's permittivity in the form
of a capacitive cell. After noise voltage has been amplified, it is
usually detected, and a DC constant of output voltage equal to a root-mean-square
value of a signal is measured. An important advantage in measuring
intrinsic noises is absence of the external electric field of high
strength that is present in the medium of the dielectric spectroscopy
technique. This field can considerably distort the results of measuring
electrical properties of the medium.

However, ferroelectric cells produce not only the thermal noises, but
also some other electric fluctuation of different nature. For example,
at frequencies less than $1\,kHz$, Barkhausen noise related to polarization
jumps can dominate if the external electric field, the temperature
and the pressure are changed \cite{Bed2011}.

According to the above mentioned research, two different approaches
in revealing nonlinearity of the moist porous medium at low frequencies
were used. The first one consisted in searching for nonlinearity
manifestations during capacity and resistance measurements at different
amplitudes of the cell test signal, as well as negative resistance
occurrence. As to the second approach, the low-frequency noise voltage
that is considerably changed in the ferroelectric, which temperature
is temporally varied, was researched. Meanwhile, we paid attention
to the registration of unsteady processes in the form of sharp voltage
pulses of short duration, similar to Barkhausen noise

An important aspect of the research was to reveal that a considerable
portion of water in the pores of this diameter had the properties
similar to volumetric water. According to some works \cite{LCh2012,FM2012}, loss
of volumetric water properties happens with pore diameter of $\sim1-2\,nm$.
The same was confirmed in \cite{CGAD2009}, when the methods of molecular dynamics
were used to study rotation and translatory motion of water molecules
in a flat layer close to hydrophilic surfaces. The surface effect
disappeared at a distance more than $0.5\,nm$.

\section*{Peculiaritied of measurement methods}

\subsection{Dielectric spectroscopy}

Measurements of the cell impedance were taken; its capacity depending
on the temperature for different amplitudes of the test sinusoida
signal in the range from $0.01\,V$ to $2\,V$ was calculated. The cell was
a plane capacitor $30\times30\,mm^2$ in size and had $4\,mm$ between the capacitor
plates. Possibility of negative resistance appearance was examined at
the same time.

The cell impedance was measured by GW-INSTEK LCR-78110G meter. The
meter was preliminarily tested to register a negative resistance.
Medium temperature was measured by means of thermocouple. Signal
recording was done by Agilent's data acquisition system. Silica MCM-41
was used as a porous medium. Its moistening thermal properties were
thoroughly studied (for example, in \cite{SKF2001}). This material was used
in \cite{Bordonskiy2011study,Bordonskiy2012study,FMBO2011} for deep water supercooling. For this research we applied
MCM-41 that was synthesized in the Institute of Chemistry and Chemical
Technology, SB RAS. Its specific internal volume was $0.8\,cm^{3}/g$ and
specific internal pore surface was about $1100\,m^{2}/g$ \cite{PK2009}. The diameter
of cylindrical pores was $3.5\,nm$. For these parameters, the temperature
of phase transition $\Delta\,T_{m}$ in case of free ice melting, i.e. when heating
after deep cooling, determined by $\Delta\,T_{m}=52/(r-t)$ formula, where r is pore
radius in nanometers, $t = 0.38\,nm$, was $-38^{\circ}\,C$ \cite{SKF2001}. When moist silicate
is cooled, the diffusion of phase transition at the temperature with
the interval about $10^{\circ}\,C$ is observed \cite{SKF2001}, so we could expect liquid
water existence close to $-45^{\circ}\,C$. Some freezing temperature drop
takes place when pore filling is not complete \cite{Johari2009}.

\subsection{Intrinsic noises within the cell and the medium}

Polarization noise measurements were taken by means of a tube
cell $10\,mm$ in diameter. Round metal electrodes were on the cylinder's
base, pore powder-like material studied was placed between them.
The cell was located in a shielded chamber where cooling nitrogen
vapors were fed. Electric noises were amplified by a two-stage
amplifier with a voltage ratio $10^{3}$. The noise signal was detected
by a linear-response detector. The constant component was filtered
by a RC circuit with 1 second time constant. Voltage measured after
the detector was equal to a root-mean-square value of the amplified
noise voltage. A frequency band limited at the high frequency by $100\,Hz$ was
formed by a low-pass filter fixed between the two stages.
Amplifier's input resistance was equal to $100\,kOhm$. A frequency
band at the low frequency was limited by an input filter equal to some
hertz. A system of information recording registered output signals
with 0.5 second interval. The cooling and heating rate was
approximately equal to $2^{\circ}\,C$ per minute.

\section*{Measurement results}

Fig. 1 shows the results of measuring cell capacity by MCM-41
depending on the medium temperature with different voltage amplitudes
on the cell. The material's gravimetric moisture in the experiment was
44\% which corresponded to 55\% of pore space filling.

\begin{figure}
\includegraphics[width=0.9\columnwidth]{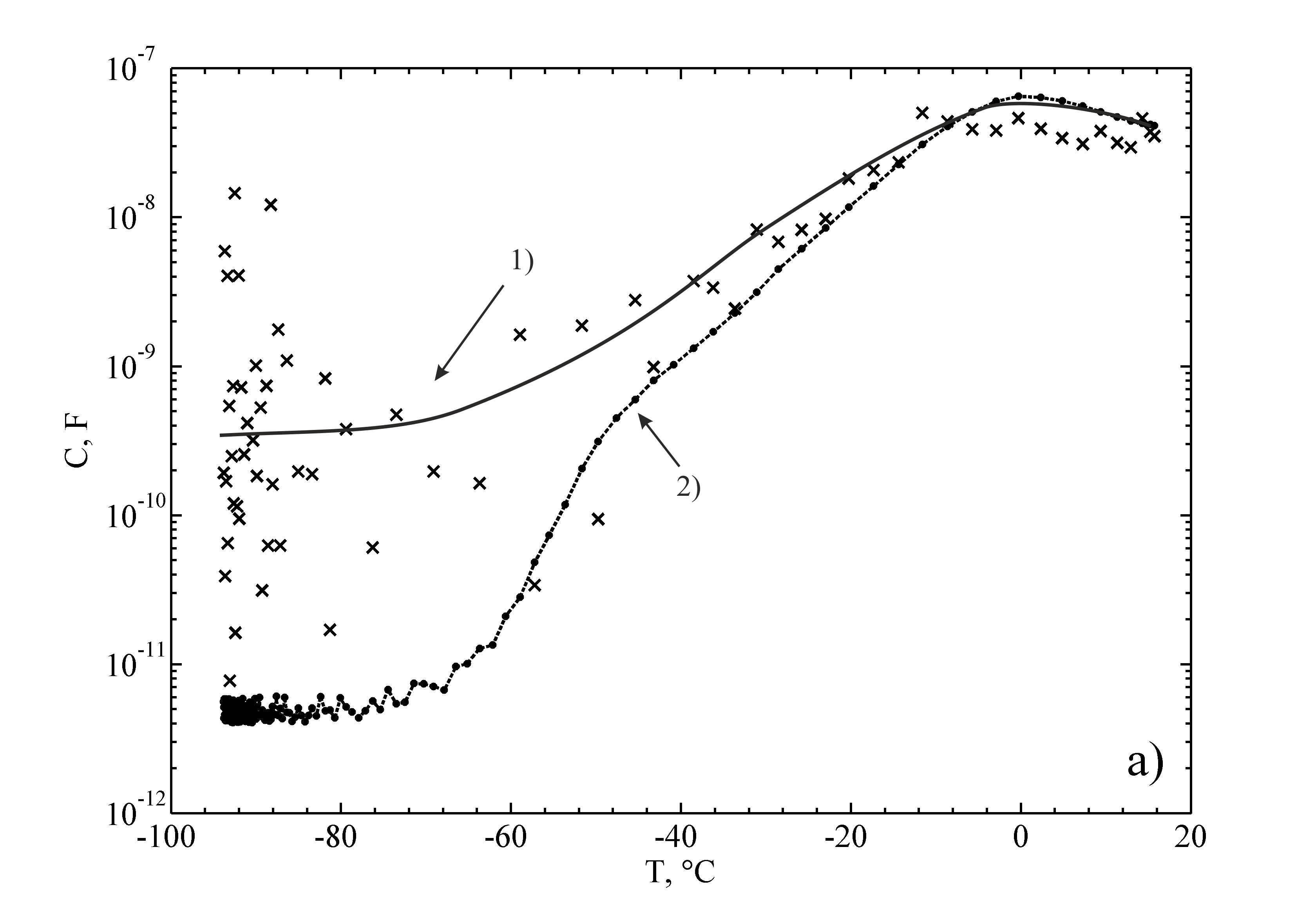}

\caption{Dependence of the measured cell capacity with moist MCM-41
sample on the temperature at different voltage amplitudes $U_{0}$ on the
cell at a frequency of $20\,Hz$, 1) $U_{0}=0.01\,V$, 2) $U_{0}=2\,V$.}
\end{figure}

There is a considerable difference in the capacity measured from the
signal's amplitude at a frequency of $20\,Hz$ at the temperature below
$-40^{\circ}\,C$. The measurements showed that for the voltage amplitude of
$0.01V$ the capacity was 100 times higher than for $0.1-2\,V$. High
capacity values at the temperatures above the phase transition
temperature are explained by the conductance effect and influence
of double electrical layer capacity at the medium and electrode
boundary \cite{BM2005}.

At frequencies about several hundred hertz and below, at measured
cell capacity dependencies on the temperature, we observed high spikes
and dips of the value measured at some temperatures (fig.1a). This
effect was observed at the temperatures below $-40^{\circ}\,C$. One of the
reasons of high instability in meter indications can be connected with
nonlinearity of an electric circuit containing the cell with the medium
studied. The reason of instability was detected by measurements of the
real part of the cell impedance.

In fig. 2 dependence of the cell impedance real part on the temperature
for the same sample is shown.

\begin{figure}
\includegraphics[width=0.65\columnwidth]{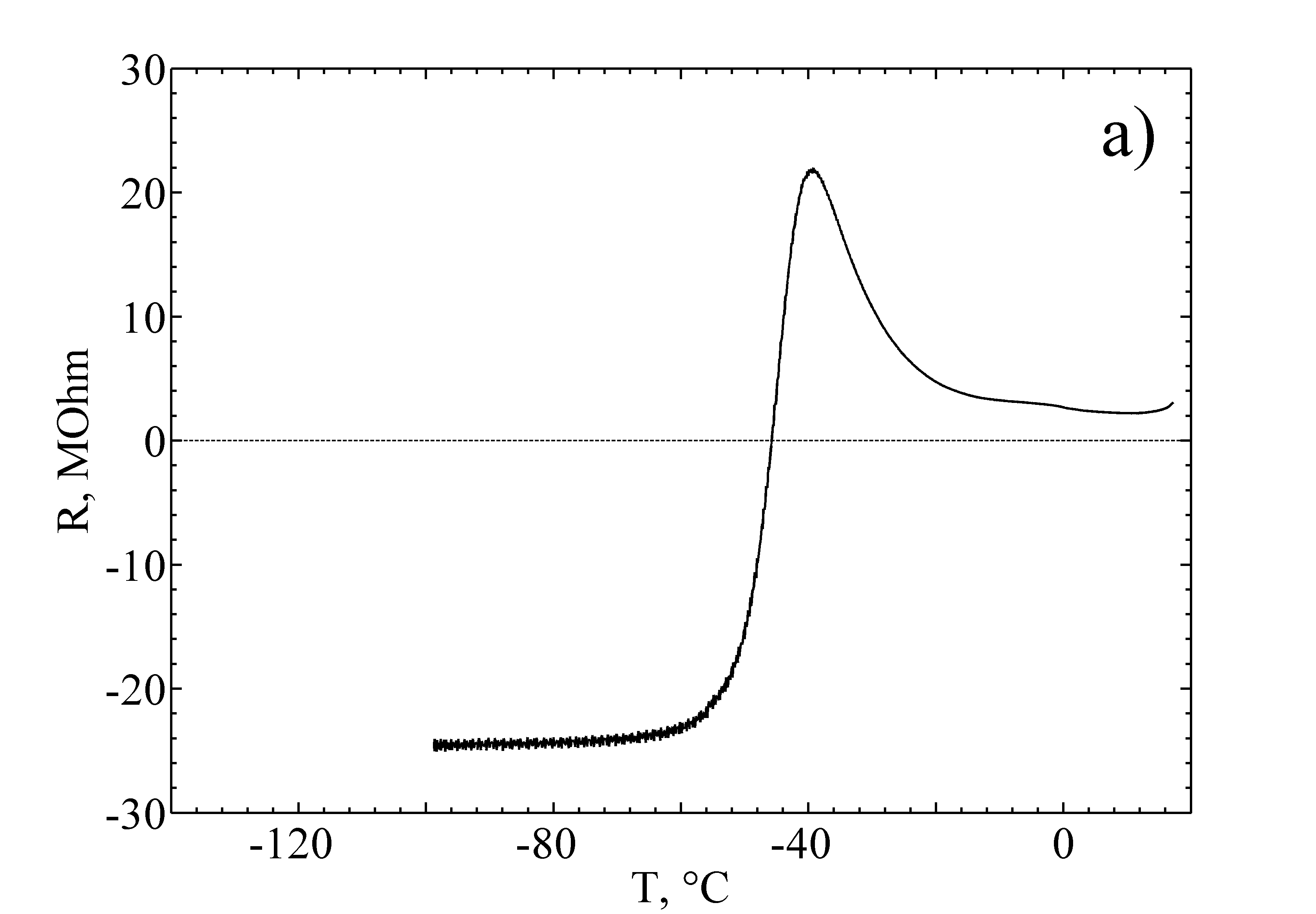}

\includegraphics[width=0.65\columnwidth]{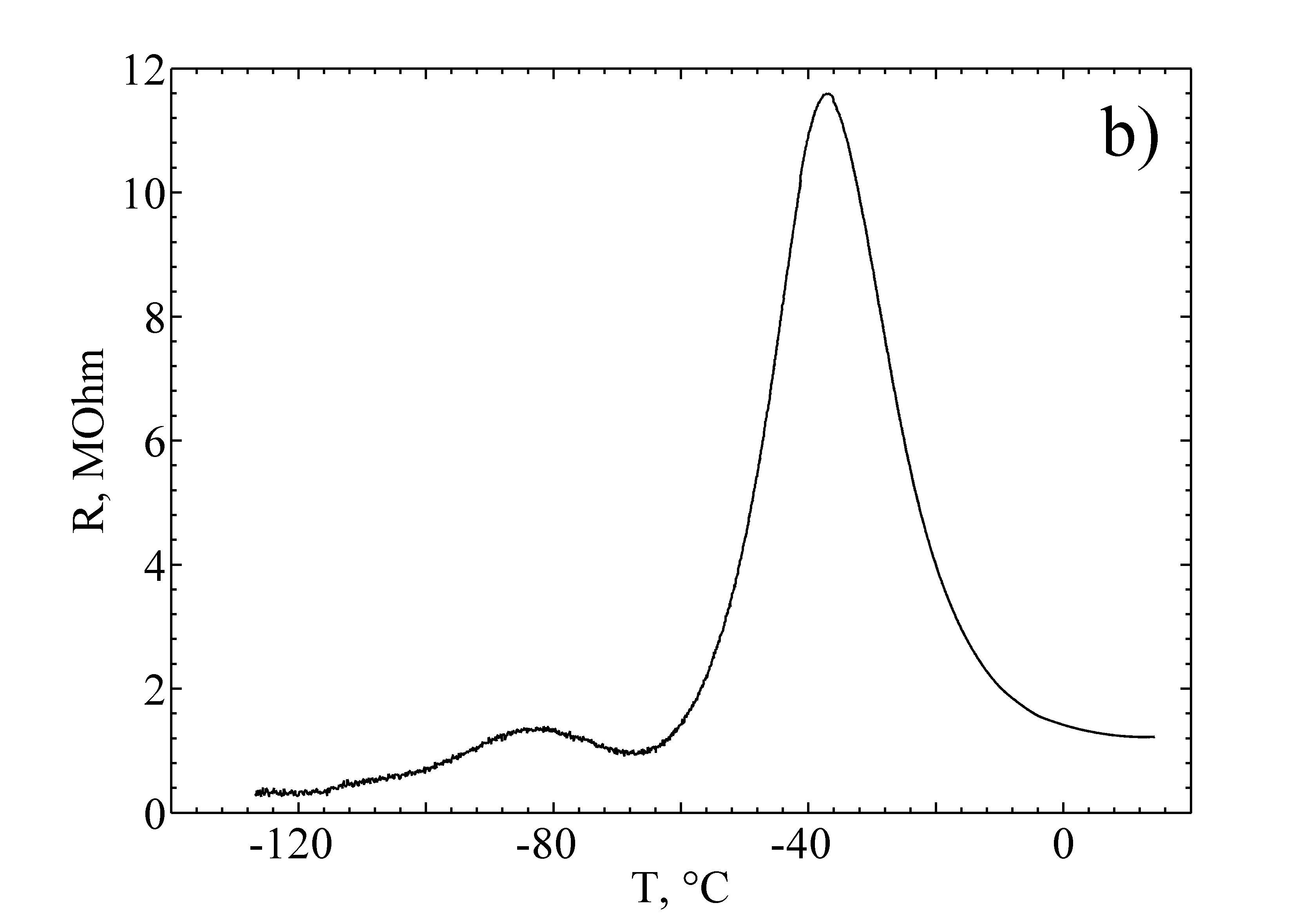}

\caption{Dependence of cell series resistance with MCM-41 on the temperature
at frequencies: (a) $20\,Hz$; (b) $1\,kHz$.}
\end{figure}

The graphs show negative resistance of the cell that appears at the
temperatures down to $-39^{\circ}\,C$ at a frequency of $20\,Hz$. There wasn't this
effect at a frequency of $1000\,Hz$. The amplitude of a test sine signal
on the cell was $0.01\,V$.

Measurements at other frequencies, in the range $20\,Hz$ to $10\,MHz$,
showed instability in the measured value during the frequency manipulation.

Taking measurements at the some frequency, one could observe spikes of the
resistance values into the negative region at frequencies about $1\,kHz$,
depending on MCM-41 sample temperature with different moisture content.
Measuring a sample with a higher phase transition temperature equal to
$-10^{\circ}\,C$ (silicate SBA-15 with $8.5\,nm$ cylinder pore diameter), there were
no negative resistance values (fig.3).

\begin{figure}
\includegraphics[width=0.65\columnwidth]{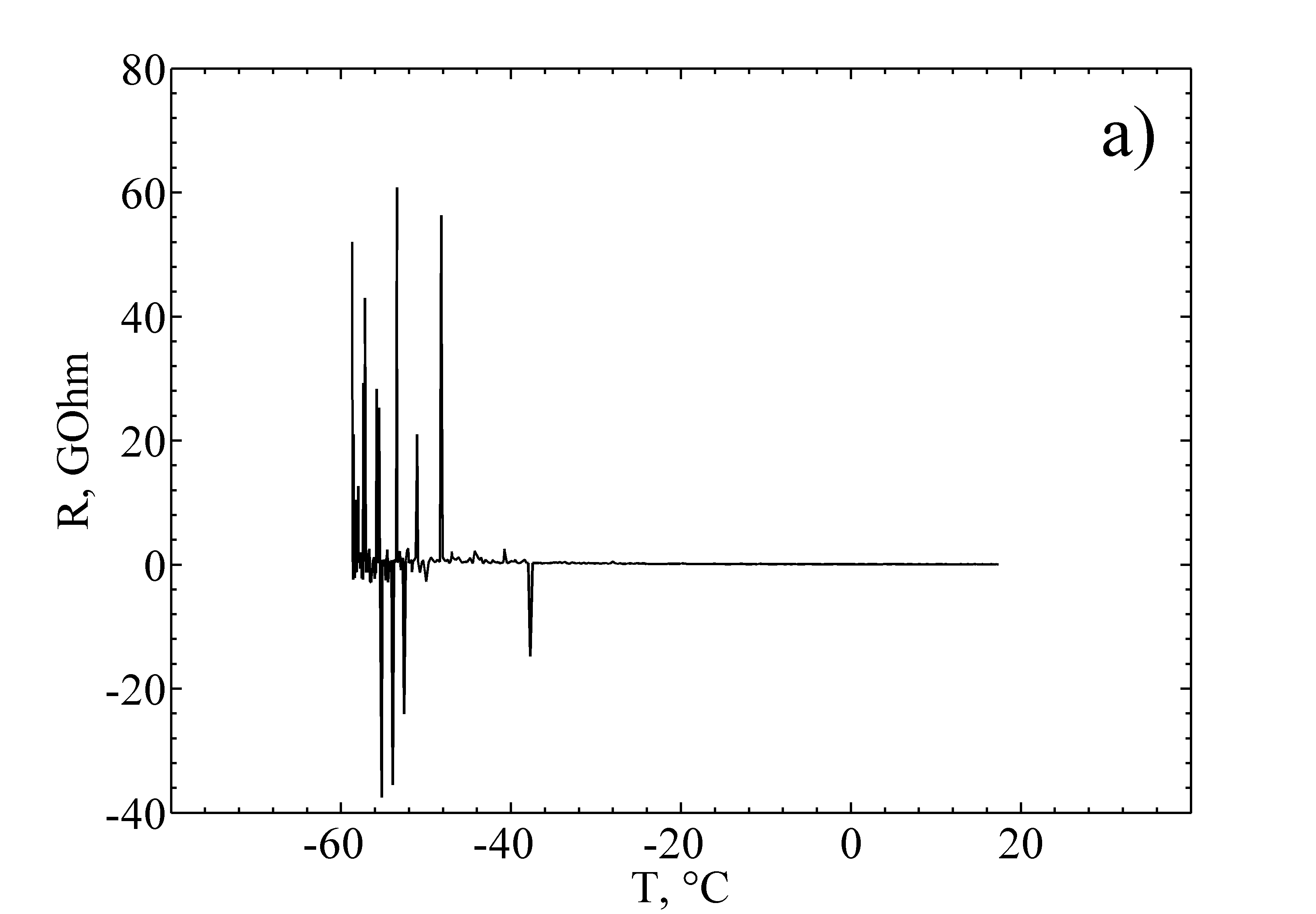}

\includegraphics[width=0.65\columnwidth]{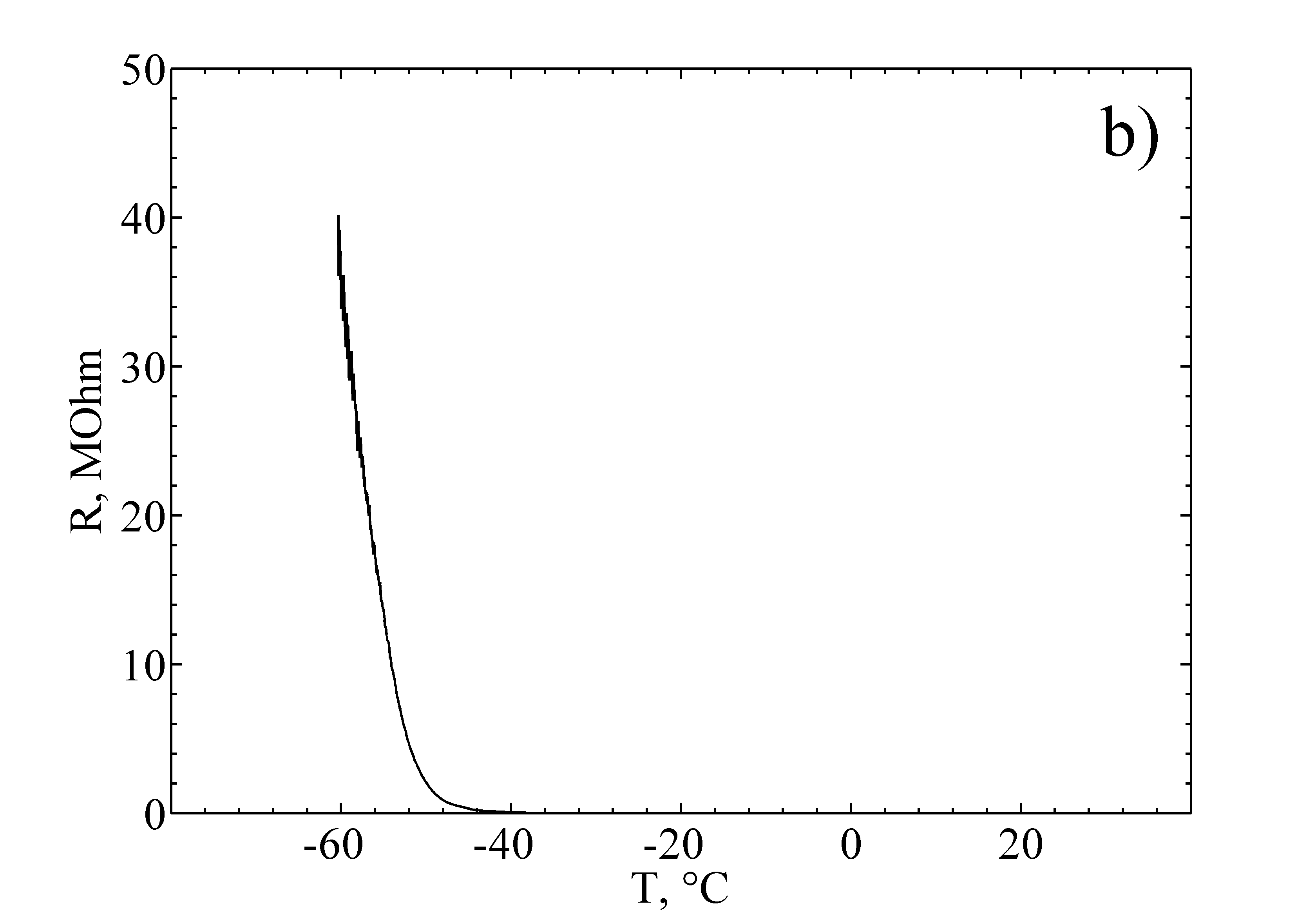}

\caption{Results of measurements on the real part of the cell impedance from
the temperature at a frequency of $1\,kHz$ for: (a) MCM-41 with a pore
diameter of $3.5\,nm$ and 15\% humidity; (b) SBA-15 with a pore diameter of
8.5 nm and 40\% humidity.}
\end{figure}

The results of measuring a root-mean-square value of electric noise amplitude
after the amplification are shown in fig. 4.

\begin{figure}
\includegraphics[width=0.9\columnwidth]{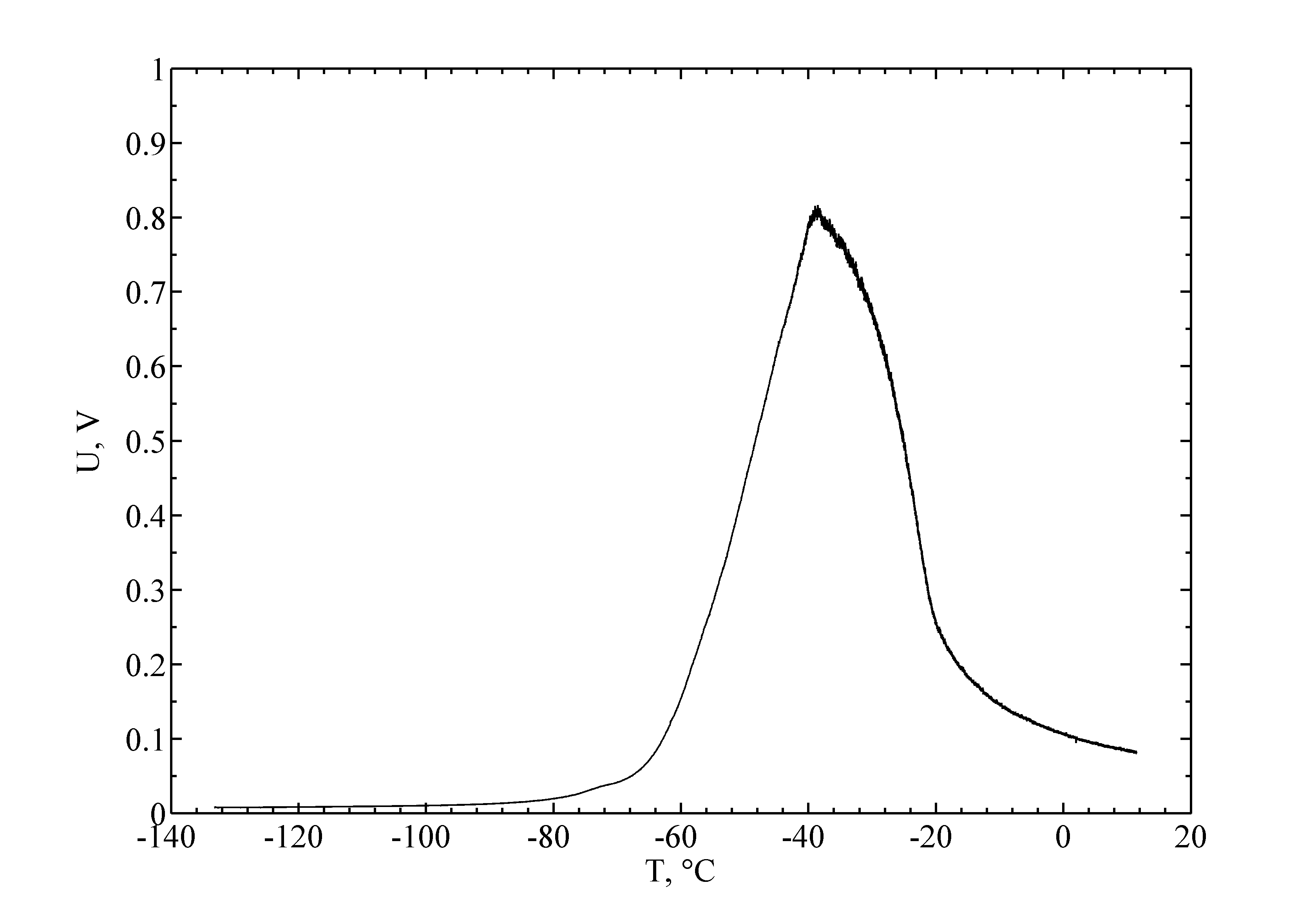}

\caption{Dependence of amplified noise EMF on the cell temperature with MCM-41
in the range $1-100\,Hz$.The sample's gravimetric moisture is 56\% corresponding
to 70\% of the pore filling.}
\end{figure}

When taking measurements for the noise voltage, a bell-shaped curve of its
dependence on the temperature with maximum $-46^{\circ}\,C$ (on medium cooling) and
$-39^{\circ}\,C$ (on its heating) was observed. Obviously, the temperature range
$-35\,\div\,-46^{\circ}\,C$ is characterized as a specific one.

Moreover, a noise EMF source can be nonstationary, as in case with Barkhausen
noise.  Data on noise signal nonstationarity can disappear during the signal
filtering and transformation. Formation of high level narrow pulses during
Barkhausen effect can be used for verification of the ferroelectric phase
existence. To detect such noise voltage pulses, we took measurements for
the electrical noise with the modified amplifier circuit. The first amplifier
stage was an operational amplifier LM833 having input bipolar
transistors; its voltage transfer ratio was equal to 5. The upper value of the
amplifier bandwidth was about $1\,MHz$ (a filter with $0-100\,Hz$ band was put
between the first and the second amplifier stages). The signal was fed to
the noninverting input. A pull-up resistor ($100\,kOhm$) was switched off
the first stage in the modified amplifier circuit and that caused a sharp
increase in input resistance. Due to leakage current in input transistor base,
the amplifier entered into the mode close to saturation which caused a decrease
in transfer coefficient for a week signal and occurrence of a high level
of direct-current voltage at the first stage circuit output.

Measuring in this mode enables to reveal the input voltage pulses about
$1\,V$ and duration up to 1 microsecond, as in this case the amplifier short-term
moves out of the saturation mode. Meanwhile, input fluctuations of the amplified
signals increase because of the transient processes in the amplifier.
The measurement results for the sample's cooling are presented
in Figure 5.  As seen from the graph of noise recording, there is a temperature
at which the noise is sharply changing. This temperature is close to
$-40^{\circ}\,C$. The same was observed using the sample heating.

\begin{figure}
\includegraphics[width=0.9\columnwidth]{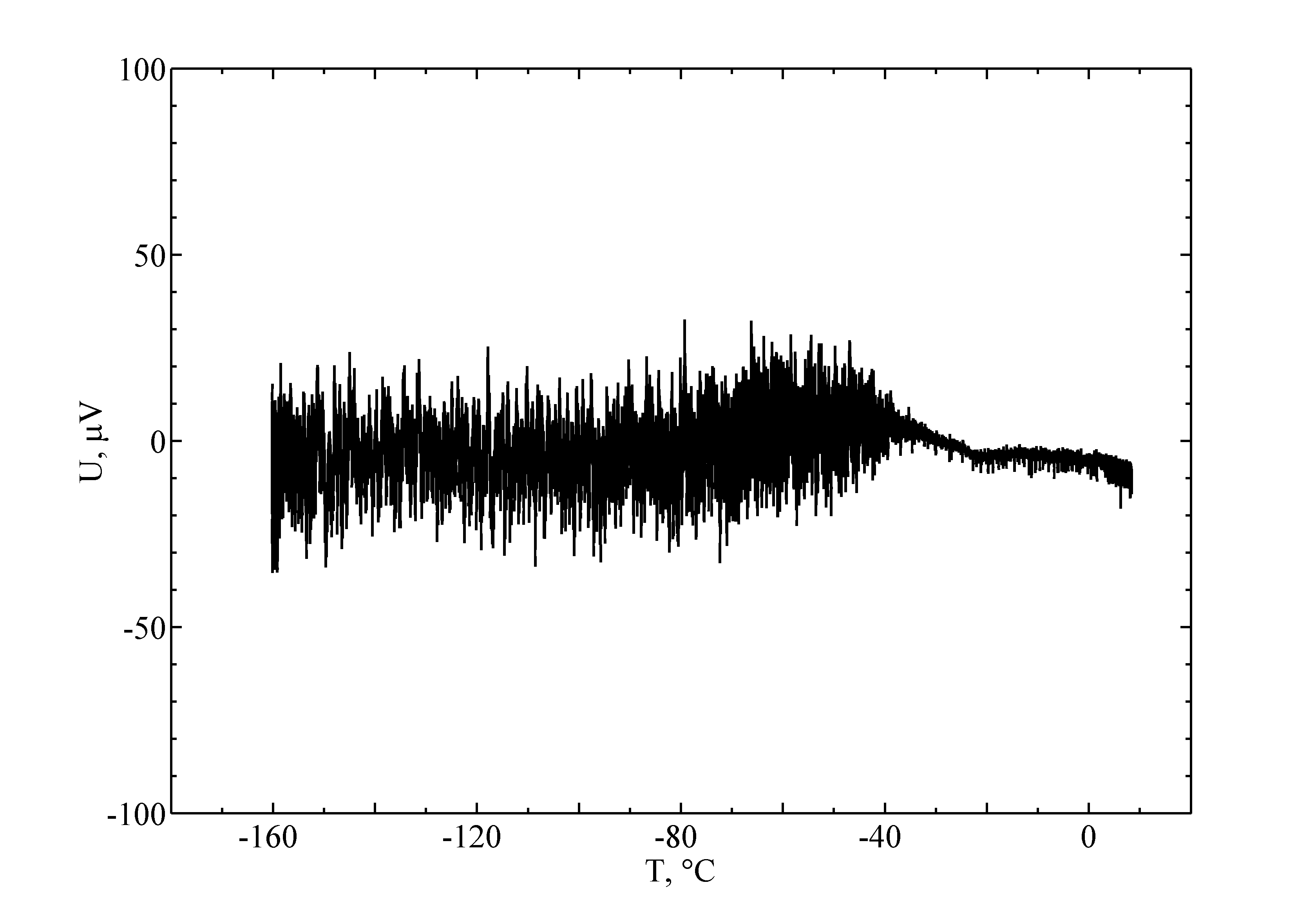}

\caption{Results of measuring cell noise with MCM-41 during the temperature change with
the amplifier in saturation mode recording high input voltage pulses, according to output
fluctuations. Cooling cycle of a sample. Gravimetric sampling moisture is 44\%.}
\end{figure}

\section*{Discution and results}

So, different ways of measuring electric properties of the moist nanoporous silicate
containing non-frozen water on moistening up to 70\% of the pore space, characterize the
temperature range close to $-40^{\circ}\,C$ as a specific one. In the temperature range
$-40\,\div\,-95^{\circ}\,C$, a 100 times capacity increase at a frequency of $20\,Hz$ with the voltage
amplitude on a cell $0.01\,V$ was observed, which corresponded to the maximum intensity
of the electric field in the medium $2.5\,V/m$. When the field intensity was higher than
$20\,V/m$, effect of the capacity increase disappeared. In a heterogeneous medium the
dielectric constant should be determined by means of Maxwell-Wagner mechanism taking
account of inter-granular polarization. However, in any case, intermittent growth
of the medium dielectric constant must be accompanied by a sharp dielectric constant
increase of one of its components which is the most probable for the pore water.
The negative resistance is registered in the medium below $-40^{\circ}\,C$. Such an effect
is feasible in nonlinear circuits; it is well-known in radio physics and is common
for different devices - vacuum-tube and transistor oscillators, Gunn-diode
oscillators, etc. Occurrence of the negative resistance
under certain conditions brings about the electric circuit instability accompanied
by the noise gain and pulse generation. The most probable reason for the
supercooled water nonlinearity is the occurrence of electric domains, i.e.
ferroelectric state.

A similar pattern of polarization noises was determined in the work \cite{MKSB1997} for the
case of ferroelectric state formation in liquid crystals on their cooling. When
passing from the positive to the negative temperatures, spectral density of the
noise voltage increased about an order of magnitude and then remained constant
in the process of aftercooling to some tens of degrees. For the measurement results
presented in Figure 4, a significant level of noises down to $-40^{\circ}\,C$ takes place.
Their increase in this case can be explained not only by Barkhausen effect, but
also by the amplification in negative resistance field determined by electric
medium instability. This effect is known in a theory of signal transformations,
for instance, for a parametric circuit in the form of paralleled capacitor and
resistor variables. Such a circuit can be found in mixer diodes, in down converter
accompanied by the noise temperature increasing
(i.e. negative resistance occurs) \cite{B1977}. In case of supercooled water parametric effect is
connected with the change in spontaneous polarization during Barkhausen effect
that can play a pumping part.

We can suppose that similarity of our results with those in the work \cite{MKSB1997}
concerning the noise occurrence after the moist medium cooling below the defined
temperature denotes the formation of ferroelectric liquid crystals in MCM-41 pores.
This hypothesis has been put forward previously \cite{Bordonskiy2011study} on the basis of both our own
results and other authors' researches \cite{WD2004,JChSch2008}. Coexistence of crystalline and
liquid states was observed by means of different methods in these works. The
hypothesis was also verified by theoretical studies \cite{BG2012}. This issue calls for
further research.

The ferroelectric phase can emerge in any nanoporous bodies due to the existence
of a number of pores less than 3 nm in size, where water supercools to about
$-40^{\circ}\,C$ and below. Moreover, liquid water films having the properties similar
to those of volumetric water, i.e. with thicknesses more than $0.5\,nm$, exist
at the interface between ice and non-freezing films of adsorbed water. For
example, supercooled water having volumetric water properties was observed
during microwave measurements of radiation power absorption through the moist
silica gel with the average pore size of $8\,nm$ at the frequencies of
$12.8-22\,GHz$ \cite{Bordonskiy2012structural}. Microwave radiation absorption was observed at the temperatures
below $-100^{\circ}\,C$ which is probable if relaxation time of water molecules is
close to oscillation period in the external electric field. The oscillation
period for these frequencies is $\sim0.1\,ns$, and that corresponds to the time
of free water relaxation.

\section*{Conclutions}

So, a complex of the electrical measurements carried out on the moist
nanoporous silicate enabling to reach deep water supercooling proves
existence of the ferroelectric phase transition at about $-40^{\circ}\,C$.

The following experimental data indicate that.

In MCM-41 at the temperature of $-38^{\circ}\,C$ we observed: a) about 100-fold
increase in dielectric conductivity; b) negative resistance occurrence
(fig. 3a); c) the noises related to Barkhausen effect (fig.5). We could
have assumed that the phenomena observed are explained by charge redistribution
occurring at the ice-liquid phase boundary during ice crystal formation,
not indicating the ferroelectric phase transition. However, this assumption
is rejected by our results. First, jump in dielectric conductivity,
Barkhausen effect in particular, definitely indicates the ferroelectric
transition. Second, if these signals had been generated by the forming
ice crystals, they would have been observed in the experiment with SBA-15
(fig. 3b) at the temperature of $-10^{\circ}\,C$, when water in cylindrical pores
$8.5\,nm$ in diameter freezes. There are no such signals, and consequently,
there are not any effects related to charge redistribution when boundaries
form. Finally, the temperature of water freezing in MCM-41 with $3.5\,nm$ pore
diameter and 50-70\% of their filling drops to $-49^{\circ}\,C$ \cite{Johari2009,JChSch2008}, i.e. water is
liquid at the temperature of $-38^{\circ}\,C$.

The research done showed that measurements for studying the ferroelectric state
should be carried out with the imposition of the external electric field with
intensity not higher than $1\,V/m$. When choosing schemes and techniques for
measuring, one should consider the negative resistance of the cell containing
the substance studied. The method of measuring intrinsic noises is effective
for studying nonlinear properties of supercooled pore water, as in this case
there is no high intensity external electric field.

Agreement of the theoretical results obtained independently \cite{MF2011} with the
experimental findings indicates that the properties of deeply supercooled pore
water can model the electrical properties of volumetric water on its supercooling,
at least at $\sim-40\,\div\,-45^{\circ}\,C$.

It is obvious that research of deeply supercooled water by conventional electric
measurement techniques (dielectric spectroscopy and thermal noise) would require
a considerable modification and correct interpretation of the data due
to occurrence of parametric phenomena in nonlinear circuits related to the
ferroelectric properties.

\section*{Acknowledgements}

This work was supported by the Russian Foundation for Basic Research,
project No. 12-05-31052.

\bibliographystyle{apsrev4-1}
\bibliography{../LGC}

\begin{thebibliography}{22}%
\makeatletter
\providecommand \@ifxundefined [1]{%
 \@ifx{#1\undefined}
}%
\providecommand \@ifnum [1]{%
 \ifnum #1\expandafter \@firstoftwo
 \else \expandafter \@secondoftwo
 \fi
}%
\providecommand \@ifx [1]{%
 \ifx #1\expandafter \@firstoftwo
 \else \expandafter \@secondoftwo
 \fi
}%
\providecommand \natexlab [1]{#1}%
\providecommand \enquote  [1]{``#1''}%
\providecommand \bibnamefont  [1]{#1}%
\providecommand \bibfnamefont [1]{#1}%
\providecommand \citenamefont [1]{#1}%
\providecommand \href@noop [0]{\@secondoftwo}%
\providecommand \href [0]{\begingroup \@sanitize@url \@href}%
\providecommand \@href[1]{\@@startlink{#1}\@@href}%
\providecommand \@@href[1]{\endgroup#1\@@endlink}%
\providecommand \@sanitize@url [0]{\catcode `\\12\catcode `\$12\catcode
  `\&12\catcode `\#12\catcode `\^12\catcode `\_12\catcode `\%12\relax}%
\providecommand \@@startlink[1]{}%
\providecommand \@@endlink[0]{}%
\providecommand \url  [0]{\begingroup\@sanitize@url \@url }%
\providecommand \@url [1]{\endgroup\@href {#1}{\urlprefix }}%
\providecommand \urlprefix  [0]{URL }%
\providecommand \Eprint [0]{\href }%
\providecommand \doibase [0]{http://dx.doi.org/}%
\providecommand \selectlanguage [0]{\@gobble}%
\providecommand \bibinfo  [0]{\@secondoftwo}%
\providecommand \bibfield  [0]{\@secondoftwo}%
\providecommand \translation [1]{[#1]}%
\providecommand \BibitemOpen [0]{}%
\providecommand \bibitemStop [0]{}%
\providecommand \bibitemNoStop [0]{.\EOS\space}%
\providecommand \EOS [0]{\spacefactor3000\relax}%
\providecommand \BibitemShut  [1]{\csname bibitem#1\endcsname}%
\let\auto@bib@innerbib\@empty
\bibitem [{\citenamefont {Frunza}\ \emph {et~al.}(2004)\citenamefont {Frunza},
  \citenamefont {Kosslick}, \citenamefont {Pitsch}, \citenamefont {Frunza},
  \citenamefont {Schonhals},\ and\ \citenamefont {et~al.}}]{Frunza2004studies}%
  \BibitemOpen
  \bibfield  {author} {\bibinfo {author} {\bibfnamefont {L.}~\bibnamefont
  {Frunza}}, \bibinfo {author} {\bibfnamefont {H.}~\bibnamefont {Kosslick}},
  \bibinfo {author} {\bibfnamefont {I.}~\bibnamefont {Pitsch}}, \bibinfo
  {author} {\bibfnamefont {S.}~\bibnamefont {Frunza}}, \bibinfo {author}
  {\bibfnamefont {A.}~\bibnamefont {Schonhals}}, \ and\ \bibinfo {author}
  {\bibnamefont {et~al.}},\ }\href@noop {} {\bibfield  {journal} {\bibinfo
  {journal} {Studies in Surface Science and Catalysis}\ }\textbf {\bibinfo
  {volume} {154}},\ \bibinfo {pages} {1721} (\bibinfo {year}
  {2004})}\BibitemShut {NoStop}%
\bibitem [{\citenamefont {Sliwinska-Bartkowia}\ \emph
  {et~al.}(2008)\citenamefont {Sliwinska-Bartkowia}, \citenamefont
  {Jazdzewska}, \citenamefont {Huang},\ and\ \citenamefont
  {Cubbins}}]{Sliwinska2008melting}%
  \BibitemOpen
  \bibfield  {author} {\bibinfo {author} {\bibfnamefont {K.}~\bibnamefont
  {Sliwinska-Bartkowia}}, \bibinfo {author} {\bibfnamefont {M.}~\bibnamefont
  {Jazdzewska}}, \bibinfo {author} {\bibfnamefont {L.}~\bibnamefont {Huang}}, \
  and\ \bibinfo {author} {\bibfnamefont {K.}~\bibnamefont {Cubbins}},\
  }\href@noop {} {\bibfield  {journal} {\bibinfo  {journal} {Phys. Chem. Chem.
  Phys.}\ }\textbf {\bibinfo {volume} {10}},\ \bibinfo {pages} {4909} (\bibinfo
  {year} {2008})}\BibitemShut {NoStop}%
\bibitem [{\citenamefont {Bordonskiy}\ and\ \citenamefont
  {Krylov}(2012)}]{Bordonskiy2012structural}%
  \BibitemOpen
  \bibfield  {author} {\bibinfo {author} {\bibfnamefont {G.}~\bibnamefont
  {Bordonskiy}}\ and\ \bibinfo {author} {\bibfnamefont {S.}~\bibnamefont
  {Krylov}},\ }\href@noop {} {\bibfield  {journal} {\bibinfo  {journal}
  {Russian Journal of Physical Chemistry A.}\ }\textbf {\bibinfo {volume}
  {86}},\ \bibinfo {pages} {1682} (\bibinfo {year} {2012})}\BibitemShut
  {NoStop}%
\bibitem [{\citenamefont {Johari}(2009)}]{Johari2009}%
  \BibitemOpen
  \bibfield  {author} {\bibinfo {author} {\bibfnamefont {G.}~\bibnamefont
  {Johari}},\ }\href@noop {} {\bibfield  {journal} {\bibinfo  {journal}
  {Thermochimica Acta}\ }\textbf {\bibinfo {volume} {492}},\ \bibinfo {pages}
  {29} (\bibinfo {year} {2009})}\BibitemShut {NoStop}%
\bibitem [{\citenamefont {Bordonskiy}\ and\ \citenamefont
  {Orlov}(2011)}]{Bordonskiy2011study}%
  \BibitemOpen
  \bibfield  {author} {\bibinfo {author} {\bibfnamefont {G.}~\bibnamefont
  {Bordonskiy}}\ and\ \bibinfo {author} {\bibfnamefont {A.}~\bibnamefont
  {Orlov}},\ }\href@noop {} {\bibfield  {journal} {\bibinfo  {journal}
  {Condensed matter and interphases}\ }\textbf {\bibinfo {volume} {13}},\
  \bibinfo {pages} {5} (\bibinfo {year} {2011})}\BibitemShut {NoStop}%
\bibitem [{\citenamefont {Bordonskiy}\ \emph {et~al.}(2012)\citenamefont
  {Bordonskiy}, \citenamefont {Gurulev}, \citenamefont {Orlov},\ and\
  \citenamefont {Schegrina}}]{Bordonskiy2012study}%
  \BibitemOpen
  \bibfield  {author} {\bibinfo {author} {\bibfnamefont {G.}~\bibnamefont
  {Bordonskiy}}, \bibinfo {author} {\bibfnamefont {A.}~\bibnamefont {Gurulev}},
  \bibinfo {author} {\bibfnamefont {A.}~\bibnamefont {Orlov}}, \ and\ \bibinfo
  {author} {\bibfnamefont {K.}~\bibnamefont {Schegrina}},\ }\href@noop {}
  {\bibfield  {journal} {\bibinfo  {journal} {Preprint arXiv: 1204.6401v1
  [cond-mat.soft]}\ ,\ \bibinfo {pages} {6}} (\bibinfo {year}
  {2012})}\BibitemShut {NoStop}%
\bibitem [{\citenamefont {Fedichev}\ \emph {et~al.}(2011)\citenamefont
  {Fedichev}, \citenamefont {Menshikov}, \citenamefont {Bordonskiy},\ and\
  \citenamefont {Orlov}}]{FMBO2011}%
  \BibitemOpen
  \bibfield  {author} {\bibinfo {author} {\bibfnamefont {P.}~\bibnamefont
  {Fedichev}}, \bibinfo {author} {\bibfnamefont {L.}~\bibnamefont {Menshikov}},
  \bibinfo {author} {\bibfnamefont {G.}~\bibnamefont {Bordonskiy}}, \ and\
  \bibinfo {author} {\bibfnamefont {A.}~\bibnamefont {Orlov}},\ }\href@noop {}
  {\bibfield  {journal} {\bibinfo  {journal} {JETP Letters}\ }\textbf {\bibinfo
  {volume} {94}},\ \bibinfo {pages} {401} (\bibinfo {year} {2011})}\BibitemShut
  {NoStop}%
\bibitem [{\citenamefont {Angell}(1983)}]{A1983}%
  \BibitemOpen
  \bibfield  {author} {\bibinfo {author} {\bibfnamefont {C.}~\bibnamefont
  {Angell}},\ }\href@noop {} {\bibfield  {journal} {\bibinfo  {journal} {Ann.
  Rev. of Phys. Chem.}\ }\textbf {\bibinfo {volume} {34}},\ \bibinfo {pages}
  {593} (\bibinfo {year} {1983})}\BibitemShut {NoStop}%
\bibitem [{\citenamefont {Menshikov}\ and\ \citenamefont
  {Fedichev}(2011)}]{MF2011}%
  \BibitemOpen
  \bibfield  {author} {\bibinfo {author} {\bibfnamefont {L.}~\bibnamefont
  {Menshikov}}\ and\ \bibinfo {author} {\bibfnamefont {P.}~\bibnamefont
  {Fedichev}},\ }\href@noop {} {\bibfield  {journal} {\bibinfo  {journal}
  {Russian Journal of Physical Chemistry A.}\ }\textbf {\bibinfo {volume}
  {85}},\ \bibinfo {pages} {906} (\bibinfo {year} {2011})}\BibitemShut
  {NoStop}%
\bibitem [{\citenamefont {Khalil}(2002)}]{Kh2002}%
  \BibitemOpen
  \bibfield  {author} {\bibinfo {author} {\bibfnamefont {H.}~\bibnamefont
  {Khalil}},\ }\href@noop {} {\emph {\bibinfo {title} {Nonlinear systems}}}\
  (\bibinfo  {publisher} {Prentice Hall},\ \bibinfo {address} {N.Y.},\ \bibinfo
  {year} {2002})\ p.\ \bibinfo {pages} {832}\BibitemShut {NoStop}%
\bibitem [{\citenamefont {Barsukov}\ and\ \citenamefont
  {Macdonald}(2005)}]{BM2005}%
  \BibitemOpen
  \bibfield  {author} {\bibinfo {author} {\bibfnamefont {E.}~\bibnamefont
  {Barsukov}}\ and\ \bibinfo {author} {\bibfnamefont {J.}~\bibnamefont
  {Macdonald}},\ }\href@noop {} {\emph {\bibinfo {title} {Impedance
  Spectroscopy. Theory, experiment and applications}}}\ (\bibinfo  {publisher}
  {John Wiley \& Sons},\ \bibinfo {year} {2005})\ p.\ \bibinfo {pages}
  {608}\BibitemShut {NoStop}%
\bibitem [{\citenamefont {Bednyakov}(2011)}]{Bed2011}%
  \BibitemOpen
  \bibfield  {author} {\bibinfo {author} {\bibfnamefont {P.}~\bibnamefont
  {Bednyakov}},\ }\emph {\bibinfo {title} {Measurements of dielectric
  properties of ferroelectric crystals and thin films by thermal noise
  methods}},\ \href@noop {} {Ph.D. thesis},\ \bibinfo  {school} {Moscow St.
  Univ.} (\bibinfo {year} {2011})\BibitemShut {NoStop}%
\bibitem [{\citenamefont {Limmer}\ and\ \citenamefont
  {Chandler}(2012)}]{LCh2012}%
  \BibitemOpen
  \bibfield  {author} {\bibinfo {author} {\bibfnamefont {D.}~\bibnamefont
  {Limmer}}\ and\ \bibinfo {author} {\bibfnamefont {D.}~\bibnamefont
  {Chandler}},\ }\href@noop {} {\bibfield  {journal} {\bibinfo  {journal} {J.
  Chem. Phys.}\ }\textbf {\bibinfo {volume} {137}},\ \bibinfo {pages} {044509}
  (\bibinfo {year} {2012})}\BibitemShut {NoStop}%
\bibitem [{\citenamefont {Fedichev}\ and\ \citenamefont
  {Menshikov}(2012)}]{FM2012}%
  \BibitemOpen
  \bibfield  {author} {\bibinfo {author} {\bibfnamefont {P.}~\bibnamefont
  {Fedichev}}\ and\ \bibinfo {author} {\bibfnamefont {L.}~\bibnamefont
  {Menshikov}},\ }\href@noop {} {\bibfield  {journal} {\bibinfo  {journal}
  {Preprint arXiv: 1206.3470 [cond-mat.soft]}\ ,\ \bibinfo {pages} {3}}
  (\bibinfo {year} {2012})}\BibitemShut {NoStop}%
\bibitem [{\citenamefont {Castrillon}\ \emph {et~al.}(2009)\citenamefont
  {Castrillon}, \citenamefont {Giovambattista}, \citenamefont {Arsay},\ and\
  \citenamefont {Debenedetti}}]{CGAD2009}%
  \BibitemOpen
  \bibfield  {author} {\bibinfo {author} {\bibfnamefont {S.-V.}\ \bibnamefont
  {Castrillon}}, \bibinfo {author} {\bibfnamefont {N.}~\bibnamefont
  {Giovambattista}}, \bibinfo {author} {\bibfnamefont {I.}~\bibnamefont
  {Arsay}}, \ and\ \bibinfo {author} {\bibfnamefont {P.}~\bibnamefont
  {Debenedetti}},\ }\href@noop {} {\bibfield  {journal} {\bibinfo  {journal}
  {J. of Phys. Chem. B.}\ }\textbf {\bibinfo {volume} {113}},\ \bibinfo {pages}
  {7973} (\bibinfo {year} {2009})}\BibitemShut {NoStop}%
\bibitem [{\citenamefont {Schreiber}\ \emph {et~al.}(2001)\citenamefont
  {Schreiber}, \citenamefont {Kotelsen},\ and\ \citenamefont
  {Findenegy}}]{SKF2001}%
  \BibitemOpen
  \bibfield  {author} {\bibinfo {author} {\bibfnamefont {A.}~\bibnamefont
  {Schreiber}}, \bibinfo {author} {\bibfnamefont {I.}~\bibnamefont {Kotelsen}},
  \ and\ \bibinfo {author} {\bibfnamefont {G.}~\bibnamefont {Findenegy}},\
  }\href@noop {} {\bibfield  {journal} {\bibinfo  {journal} {Phys. Chem. Chem.
  Phys.}\ }\textbf {\bibinfo {volume} {3}},\ \bibinfo {pages} {1185} (\bibinfo
  {year} {2001})}\BibitemShut {NoStop}%
\bibitem [{\citenamefont {Parfenov}\ and\ \citenamefont
  {Kirik}(2009)}]{PK2009}%
  \BibitemOpen
  \bibfield  {author} {\bibinfo {author} {\bibfnamefont {V.}~\bibnamefont
  {Parfenov}}\ and\ \bibinfo {author} {\bibfnamefont {S.}~\bibnamefont
  {Kirik}},\ }in\ \href@noop {} {\emph {\bibinfo {booktitle} {IV Staverov
  Reading}}}\ (\bibinfo {organization} {Krasnoyarsk Univ.},\ \bibinfo {year}
  {2009})\ p.\ \bibinfo {pages} {319}\BibitemShut {NoStop}%
\bibitem [{\citenamefont {Musevic}\ \emph {et~al.}(1997)\citenamefont
  {Musevic}, \citenamefont {Kityk}, \citenamefont {Skarabot},\ and\
  \citenamefont {Blinc}}]{MKSB1997}%
  \BibitemOpen
  \bibfield  {author} {\bibinfo {author} {\bibfnamefont {I.}~\bibnamefont
  {Musevic}}, \bibinfo {author} {\bibfnamefont {A.}~\bibnamefont {Kityk}},
  \bibinfo {author} {\bibfnamefont {M.}~\bibnamefont {Skarabot}}, \ and\
  \bibinfo {author} {\bibfnamefont {R.}~\bibnamefont {Blinc}},\ }\href@noop {}
  {\bibfield  {journal} {\bibinfo  {journal} {Phys. Rev. Lett.}\ }\textbf
  {\bibinfo {volume} {79}},\ \bibinfo {pages} {1062} (\bibinfo {year}
  {1997})}\BibitemShut {NoStop}%
\bibitem [{\citenamefont {Bordonskiy}(1977)}]{B1977}%
  \BibitemOpen
  \bibfield  {author} {\bibinfo {author} {\bibfnamefont {G.}~\bibnamefont
  {Bordonskiy}},\ }\href@noop {} {\emph {\bibinfo {title} {Questions of the
  analysis of millimeter-wave frequency converters on diodes with a Schottky
  barrier}}},\ \bibinfo {type} {Tech. Rep.}\ \bibinfo {number} {TM 75049}\
  (\bibinfo  {institution} {NASA Technical Reports},\ \bibinfo {year}
  {1977})\BibitemShut {NoStop}%
\bibitem [{\citenamefont {Webber}\ and\ \citenamefont {Dore}(2004)}]{WD2004}%
  \BibitemOpen
  \bibfield  {author} {\bibinfo {author} {\bibfnamefont {B.}~\bibnamefont
  {Webber}}\ and\ \bibinfo {author} {\bibfnamefont {J.}~\bibnamefont {Dore}},\
  }\href@noop {} {\bibfield  {journal} {\bibinfo  {journal} {J. of Physics:
  Condens. Matter}\ }\textbf {\bibinfo {volume} {16}},\ \bibinfo {pages} {5449}
  (\bibinfo {year} {2004})}\BibitemShut {NoStop}%
\bibitem [{\citenamefont {Jahnert}\ \emph {et~al.}(2008)\citenamefont
  {Jahnert}, \citenamefont {Chaves}, \citenamefont {Schaumann},\ and\
  \citenamefont {et~al.}}]{JChSch2008}%
  \BibitemOpen
  \bibfield  {author} {\bibinfo {author} {\bibfnamefont {S.}~\bibnamefont
  {Jahnert}}, \bibinfo {author} {\bibfnamefont {F.}~\bibnamefont {Chaves}},
  \bibinfo {author} {\bibfnamefont {G.}~\bibnamefont {Schaumann}}, \ and\
  \bibinfo {author} {\bibnamefont {et~al.}},\ }\href@noop {} {\bibfield
  {journal} {\bibinfo  {journal} {Phys. Chem. Phys.}\ }\textbf {\bibinfo
  {volume} {10}},\ \bibinfo {pages} {6039} (\bibinfo {year}
  {2008})}\BibitemShut {NoStop}%
\bibitem [{\citenamefont {Blinov}\ and\ \citenamefont {Golo}(2012)}]{BG2012}%
  \BibitemOpen
  \bibfield  {author} {\bibinfo {author} {\bibfnamefont {V.}~\bibnamefont
  {Blinov}}\ and\ \bibinfo {author} {\bibfnamefont {V.}~\bibnamefont {Golo}},\
  }\href@noop {} {\bibfield  {journal} {\bibinfo  {journal} {JETP Letters}\
  }\textbf {\bibinfo {volume} {96}},\ \bibinfo {pages} {475} (\bibinfo {year}
  {2012})}\BibitemShut {NoStop}%
\end{thebibliography}%

\end{document}